\documentstyle[12pt]{article}
\begin{document}
\begin {flushright}
ITP-SB-95-60\\ LBL-38282
\end {flushright} 
\vskip 0.3cm
\begin{center}
{\Large \bf Charm and Bottom Quark Production Cross Sections Near
Threshold} 
\end{center}
\vskip 0.3cm \centerline{\sc J. Smith} 
\vskip 0.3cm \centerline{\it Institute for Theoretical Physics,}
\centerline{\it State University of New York at Stony Brook,}
\centerline{\it Stony Brook, NY 11794-3840}   
\vskip 0.3cm \centerline{\sc R. Vogt\footnote{This work was supported
in part by the Director, Office of Energy Research, Division of
Nuclear Physics of the Office of High Energy and Nuclear Physics of
the U. S.  Department of Energy under Contract Number
DE-AC03-76SF0098.}} 
\vskip 0.3cm \centerline{\it Nuclear Science Division,}
\centerline{\it Lawrence Berkeley National Laboratory,}
\centerline{\it Berkeley, California 94720} \centerline{\it and}
\centerline{\it Physics Department,} \centerline{\it University of
California at Davis,} \centerline{\it Davis, California 95616 USA}
\vskip 0.3cm

\begin{center}
April 1996
\end{center}

\begin{abstract}
The cross sections for charm and bottom quark production  in the
threshold region are discussed.  We consider the effects of an all
order resummation of initial state soft-plus-virtual gluon radiation
on the total cross sections compared to the order $\alpha_s^3$
results. 
\end{abstract}

\pagebreak
Heavy quark production has long been a topic of interest, both
experimentally and theoretically.  Early measurements of the total $c
\overline c$ production cross section  at $\sqrt{S} \leq 63$ GeV
suggested that the calculated Born (LO) cross section underpredicted
the data by a factor of two to three \cite{Reu,Appel},  known as the
$K$ factor after a similar situation in Drell-Yan production.  In
general,
\begin{equation} K_{\rm exp} = \frac{\sigma_{\rm data}(AB 
\rightarrow Q \overline Q)}{\sigma_{\rm theory}(AB \rightarrow Q \overline Q)} 
\, \, , \end{equation} where $Q$ is the heavy quark ($c$, $b$, or $t$)
and $\sigma_{\rm
theory}$ is calculated to fixed order in the running coupling constant
$\alpha_s$.  The next-to-leading order (NLO) 
corrections to the Born cross section have
been calculated \cite{nde1,betal,bnmss} and an analogous 
theoretical $K$ factor is
defined by the ratio of the NLO to the LO cross sections, \begin{equation}
K_{\rm th} = \frac{\sigma^{\rm NLO}(AB 
\rightarrow Q \overline Q)}{\sigma^{(0)}(AB \rightarrow Q \overline Q)} 
\, \, , \end{equation} where $\sigma^{(0)}$ is the LO cross section and
$\sigma^{\rm NLO}$ is the sum of the LO and the exact
${\cal O}(\alpha_s)$ correction, $\sigma^{\rm NLO} = \sigma^{(0)} + 
\sigma^{(1)}\mid _{\rm exact}$.

The heavy quark production cross section is calculated
in QCD by assuming the validity of the factorization theorem \cite{css} and 
expanding the contributions to the amplitude in powers of the coupling 
constant $\alpha_s(\mu^2)$. 
The hadronic production cross section at hadronic center of mass energy
$\sqrt{S}$ and to order $\alpha_s^{(k)}$ is
\begin{equation}
\sigma^{(k)}(S,m^2) =  \sum_{ij} \int_{\frac{4m^2}{S}}^1 d\tau
\int_\tau^1 \frac{dx}{x}
f_i^{h_1}(x,\mu^2) f_j^{h_2}(\frac{\tau}{x},\mu^2) \sigma_{ij}^{(k)}(\tau S, 
m^2,\mu^2),  
\end{equation}
where $f_i^h(x,\mu^2)$ are the scale-dependent parton densities of hadron $h$
evaluated at the
scale $\mu^2$ and $\sigma_{ij}^{(k)}$ is the $k^{\rm th}$ order
partonic cross section for $ij \rightarrow Q \overline Q X$. 
The total cross section up to  order $k$ is $\sum_k
\sigma^{(k)}(S,m^2)$.
The numerical results for ``lighter'' heavy quark production
depend on the choice of the
parton densities, involving the mass factorization
scale $\mu_F$, the running coupling constant, evaluated at
the renormalization scale $\mu_R$, and the heavy quark mass $m$.  Usually 
the renormalization and factorization scales are assumed to be equal, $\mu_R 
= \mu_F = \mu$.  All these factors influence
both $K_{\rm exp}$ and $K_{\rm th}$.  

At LO the cross section is very sensitive to
the mass and scale parameters. However even including the NLO
corrections cannot completely fix the cross section. The sensitivity
to even higher terms in the QCD expansion is often
demonstrated by varying $\mu$ between $m/2$ and $2m$.  This may not be
very meaningful, especially for charm, since a variation of
an order of magnitude or more is observed.  It is therefore 
not clear that the next-to-next-to-leading order (NNLO) corrections are
not at least as large as the NLO corrections, particularly
when $m \ll \sqrt{S}$.   Given these facts,
it is impossible to make more precise
predictions in the absence of a NNLO calculation.  

Recently two groups have attempted to use the NLO calculations to make more
definitive statments about charm production \cite{HPC,RV,MLM1}. The NLO 
calculations were compared to charm production 
data in $pp$ and $\pi^-
p$ interactions \cite{Reu,Appel,proton,pion}.
In the first approach \cite{HPC,RV}, 
the data were used to fix $m_c$ and $\mu$ by requiring
$K_{\rm exp}^{\rm NLO} \sim 1$ in an attempt to place bounds on
charm production at nuclear colliders.  Two recent parton 
densities\footnote{All available 
parton densities for the nucleon and the pion can be found in PDFLIB 
\cite{PDFLIB}.},
MRS D$-^\prime$ \cite{SMRS,mrs} and  GRV HO \cite{GRVpi,GRV} were used in the
comparison.  Reasonable
agreement was found for $m_c = 1.2$ GeV$/c^2$, $\mu = 2m_c$ with 
MRS D$-^\prime$ \cite{SMRS,mrs}
and for $m_c = 1.3$ GeV$/c^2$, $\mu = m_c$ with GRV HO
\cite{GRVpi,GRV} although both results tend to 
underestimate the total $c \overline c$ production cross section,
$\sigma_{c \overline c}^{\rm tot}$, with $K_{\rm exp}^{\rm NLO} 
\sim 1.1 - 2$.  In the
range of the parameter space defined by $m_c$, $\mu_R$ and $\mu_F$, 
$K_{\rm exp}^{\rm NLO}$ can be reduced to unity.  
However, it is questionable if
the mass and scale values needed for $K_{\rm exp}^{\rm NLO} \sim 1$ 
are consistent with a perturbative treatment
and with the defined limits of the parton density 
distributions.  In another approach, calculations using a charm mass of 
$m_c =1.5$ GeV$/c^2$ produced results compatible with the data although
with some essential caveats: 
$\mu_F$ and $\mu_R$ were varied independently and out-of-date
parton distributions fit with several values of
$\Lambda_{\rm QCD}$ were used \cite{MLM1}.  
Decreasing $\mu_R$ with respect to $\mu_F$ and
increasing $\Lambda_{\rm QCD}$ result in significantly larger cross
sections for a given $m_c$.  Additionally, different parton densities were used
in the calculations of $\sigma_{Q \overline Q}^{\rm tot}$ and high energy $b$
production.  Neither approach is fully satisfactory:  either an uncomfortably
small charm quark mass is needed or the parameters used to describe
low energy production are incompatible with those used at collider
energies.  

Although a complete calculation of still higher order terms is not possible 
for all values of $S$ and $m$, improvements may 
be made in specific kinematical regions.
Investigations have
shown that near threshold there can be large logarithms in the perturbative
expansion which must be resummed to make more reliable theoretical
predictions. These large logarithms arise 
from an imperfect cancellation
of the soft-plus-virtual (S+V) terms.
In \cite{LSN} an approximation was given for the S+V gluon contributions
and the analogy with the Drell-Yan process, studied in \cite{gs,CT}, 
was exploited to resum these to all orders of perturbation theory. 
The same resummation procedure was also applied to $\sigma_{t \overline t}^{\rm
tot}$ and inclusive top quark distributions in $p \overline p$ collisions
at the Fermilab Tevatron \cite{LSN,lsn2,NKJS}.  For top
production in $p \overline p$ collisions, $q \overline q$ annihilation is the
dominant process, fortunate since the exponentiation of the S+V terms 
\cite{LSN} is better understood because the simple color structure has a close 
correspondence to the Drell-Yan studies \cite{gs,CT}.

The resummation of the leading S+V terms \cite{LSN} modifies eq.\ (3) so that
\begin{eqnarray}
\sigma^{\rm res}(S,m^2) =  \sum_{ij} \int_{\tau_0}^1 d\tau
\int_\tau^1 \frac{dx}{x}
f_i^{h_1}(x,\mu^2) f_j^{h_2}(\frac{\tau}{x},\mu^2)
\sigma_{ij}^{\rm res}(\tau S, m^2),
\end{eqnarray}
where
\begin{eqnarray}  
\sigma_{ij}^{\rm res}(\tau S, m^2) = -\int_{s_0}^{s-2ms^{1/2}} ds_4
f(\frac{s_4}{m^2},\frac{m^2}{\mu^2})
\frac{d\overline \sigma_{ij}^{(0)}(s,s_4,m^2)}{ds_4}.
\end{eqnarray} 
The integration variable $s_4$, an invariant which measures the four-momentum
carried away by the final-state gluon in the process $i(k_1) + j(k_2) 
\rightarrow Q(p_1) + \overline Q(p_2) + g(k_3)$, is defined as
$s_4 = s + t + u - 2m^2$
where $s = (k_1 + k_2)^2$, $t = (k_1 - p_1)^2$, and $u = (k_1 - p_2)^2$.  The
cross section $\overline \sigma_{ij}^{(0)}(s,s_4,m^2)$ is the angle-averaged
Born cross section and  $\overline \sigma_{ij}^{(0)}(s,0,m^2) \equiv 
\sigma_{ij}^{(0)}(s,m^2)$.  The function $f({s_4}/{m^2},{m^2}/{\mu^2})$ is
\begin{equation}
f\left(\frac{s_4}{m^2},\frac{m^2}{\mu^2}\right)=
\exp\left[A\frac{C_{ij}}{\pi}\overline\alpha_s\left(\frac{s_4}{m^2},m^2\right)
\ln^2\frac{s_4}{m^2}\right]
\frac{[s_4/m^2]^{\eta}}{\Gamma(1+\eta)}\exp(-\eta \gamma_E)\,,
\end{equation}
where $\gamma_E$ is the Euler constant, $\eta = (8C_{ij}/\beta_0)\ln(1 +
({\beta_0}\alpha_s(\mu^2)/4\pi) \ln(m^2/\mu^2))$, $\beta_0 = 11 - 2n_f/3$
for SU(3), and $A$ and $\overline\alpha_s$ are scheme dependent.  In 
{\it e.g.}\ the 
$\overline{\rm MS}$ scheme, $A=2$ and $\overline\alpha_s((s_4/m^2),m^2)
= \alpha_s(s_4^{2/3}m^{2/3})$.  The integral over $s_4$ has been split into
two parts and the integral over the region $0<s_4<s_0$, where the running 
coupling becomes infinite as $s_4 \rightarrow 0$, is assumed to be negligible.
This condition is satisfied if $s_0/m^2 \ll 1$.  The lower limit on the $\tau$
integral is
\begin{equation}
\tau_0 = \frac{[m+(m^2+s_0)^{1/2}]^2}{S}\,,
\end{equation}
where $s_0=m^2(\mu_0^2/\mu^2)^{3/2}$ in the $\overline{\rm MS}$ scheme and
$s_0=m^2(\mu_0^2/\mu^2)$ in the DIS scheme. The cutoff parameter $\mu_0$, 
introduced in \cite{CT,AMS}, monitors the sensitivity of the cross section to
nonperturbative higher-twist effects.  If $\sigma_{ij}^{\rm res}$ is strongly
dependent on $\mu_0$, a precise determination of the cross section requires
full knowledge of the nonperturbative contributions.  As may be expected, the
resummed corrections diverge for small $\mu_0$ \cite{LSN}.

The resummation technique has recently been applied to $b \overline b$ 
production in $pp$ interactions \cite{KS1} at HERA-B ($\sqrt{S} = 39.2$ GeV)
\cite{herab1,herab2} where gluon fusion is the dominant channel.
The $gg$ channel is not as amenable to the resummation technique because each
of the Born diagrams has a different color structure\footnote{Quark-gluon 
scattering, producing a final state quark or
antiquark, cannot be resummed by this method 
since there is no equivalent Born term.}.  
Therefore it is important to test the model in
a regime where data already exists. Charm and bottom production 
data from $pp$ and $\pi^- p$ interactions
are available at energies where the resummation technique can be applied.
An advantage of the pion beam is
the relative importance of the $q \overline q$ channel in $c \overline c$ and
$b \overline b$ production.
However charm production must be treated with some care.  The data is
available for $\sqrt{S} \geq 15$ GeV where the expansion parameter,
$\alpha_s(m^2) \ln(\sqrt{S}/m)$, is not small.  However, if the model is
reliable for charm production, a better understanding of $c \overline c$
production at the lower fixed-target energies may be reached.

Since we wish to compare our results with data taken using both pion and proton
beams, we are rather constrained in our choice of parton distribution
functions.  The only consistent NLO evaluation of the pion and proton
parton densities has been done by GRV \cite{GRVpi,GRV}.  Because the
GRV HO parameterization is in the $\overline{\rm MS}$ scheme, 
we are forced
to use this scheme for both the $q \overline q$ and $gg$ channels.  Thus the
value of $\mu_0$ needed for stability of the resummed result in the $q
\overline q$ channel is likely to be larger than found previously
\cite{LSN,lsn2,NKJS,KS1}.   
Our results for the exact, approximate, and resummed
hadronic cross sections will be calculated using
these distributions along with the two-loop
uncorrected running coupling constant\footnote{The difference between the
uncorrected running coupling constant and the corrected value given by PDFLIB
\cite{PDFLIB} is small, $\approx 3$-4\%.}.  This set has the
additional advantage of a rather small initial scale so that we can use
$\mu = m_c$ in our calculations.  We therefore take 
$m_c = \mu = 1.5$ GeV$/c^2$ and
$m_b = \mu = 4.75$ GeV$/c^2$, along with the GRV HO parton densities and the
$\overline{\rm MS}$ scheme for our principle results.  We will also show the
range of resummed cross sections by varying all these parameters, including
the parton densities.  In Table 1
we give the value of $\Lambda_{n_f}$, $\mu$ and $\alpha_s$ for each value of
$m$ and set of parton densities we consider.  The choice of these particular
parameters will be explained later.

We first review the NLO contributions to $c \overline c$ and $b \overline b$
production \cite{nde1,betal,bnmss}.
In Fig.\ 1 we show the relative contributions of the $gg$, $q \overline q$, and
the $(q+\overline q)g$ channels as functions of $\sqrt{S}$ for $c
\overline c$ and $b \overline b$ production in $\pi^- p$ and $pp$ interactions.
It is clear that $c \overline c$ production is almost entirely dominated by
gluon fusion. In $\pi^- p$ interactions at $\sqrt{S} \approx 10$ GeV the $gg$
channel accounts for half the cross section.  However, the gluon contribution
increases to more than 80\% soon afterward and the $gg$ channel in 
$pp$ production is always $\approx 90$\% of the total cross section. Note that
the quark-gluon scattering contribution is negative at low energies but changes
sign at $\sqrt{S} \approx 40$ GeV.  The $q \overline q$ annhilation
contribution is larger for $b \overline b$ production, particularly for $\pi^-
p$ interactions where this channel is dominant until $\sqrt{S} \approx 40$ GeV.
This situation is similar to top production in $p \overline p$ collisions at
the Fermilab Tevatron.  The $pp$ production cross section is again dominated by
gluon fusion, between 70-80\% of the total, somewhat less than the $gg$
contribution to $c \overline c$ production at the same energies.  
Quark-gluon scattering
always gives a negative contribution to the total $b \overline b$
cross section for the energies considered.  
Note that although we show the $b\overline b$ results for
$\sqrt{S} = 10$ GeV, the calculation is most reliable for $\sqrt{S} \geq 20$
GeV. 

The resulting theoretical $K$ factors for the total cross sections reflect
the uncertainties in the dominant channels.
Thus for $c \overline c$ production, $K_{\rm th}>2$, near the
gluon $K_{\rm th}$, over a large energy range.  Likewise, $K_{\rm th}$ is
smaller for $\pi^- p \rightarrow b \overline b$ interactions because of the
dominant $q \overline q$ contribution for $\sqrt{S}\leq 40$ GeV.
The $q \overline q$ 
channel is most amenable to resummation because even above threshold the NLO
corrections are small.  Above threshold in the $gg$ channel, the NLO
contribution increases as a function of $\sqrt{S}/2m$ and becomes constant at
high energies while the Born
contribution decreases to zero, resulting in large corrections. 
In $b \overline b$ production $K_{\rm th}$ is large 
for both channels near threshold.
 
In Fig.\ 2 we examine the $\mu_0$ dependence of the 
resummed cross section using our parameters
for $c \overline c$ production at $\sqrt{S} = 15$ GeV and 
$b \overline b$ production at $\sqrt{S} = 30$ GeV.
We also show, for comparison, the $\mu_0$ dependence of the approximate NNLO
cross section,
\begin{equation}
\sigma^{\rm app} \equiv 
\sigma^{(0)}+\sigma^{(1)}\mid _{\rm app}+\sigma^{(2)}\mid _{\rm app} \,,
\end{equation}
where we have imposed the 
same $s_4$ ($s_4>s_0$) phase space cut on $\sigma^{\rm app}$
as on the resummed cross section. Here $\sigma^{(1)}\mid _{\rm app}$
and $\sigma^{(2)}\mid _{\rm app}$ denote the approximate first and second
order corrections, respectively, where only soft gluon contributions
are taken into account.  Note that the phase space cut is also applied to
$\sigma^{(0)}$. The effect of the resummation is apparent
in the difference between the curves.
At small $\mu_0$, $\sigma^{\rm res}$ diverges, signalling the 
onset of nonperturbative physics.
In a study of $b \overline b$ production
at HERA-B \cite{KS1}, $\mu_0$ was chosen so that the resummed cross section 
remains somewhat larger 
than $\sigma^{\rm app}$ in each channel, namely 
$\mu_0 = 0.13m$ in the $q \overline q$ channel with
the MRS D$-^\prime$ DIS distributions and $\mu_0 = 0.36m$ in
the $gg$ channel, quite similar to earlier choices for top quark production
\cite{lsn2}\footnote{An application of principal value 
resummation leads to very similar results for the top quark production cross 
section \cite{BC}.}.  We choose the same ratios
\cite{KS1} for the $gg$ channel, so that $\mu_0
\approx 0.35m$.  Reasonable convergence is seen for this value, even for $c
\overline c$ production, although $\sigma^{\rm res}$ remains larger than
$\sigma^{\rm app}$ for all values of $\mu_0/m$ shown.  Even though
our calculation is in
the $\overline{\rm MS}$ scheme, convergence is found for $\mu_0
\approx 0.15m$ in the $q \overline q$ channel, not significantly larger than
that of the DIS scheme. 

In Fig.\ 3 we plot the resummed
cross section with our chosen values of $\mu_0$
as a function of center of mass energy for $c \overline c$ and $b \overline b$
production at fixed-target energies.
Since the exact $O(\alpha_s^3)$ results are known, we also show
the perturbation theory improved cross sections
defined by
\begin {equation}
\sigma^{\rm imp}=\sigma^{\rm res}
+\sigma^{(1)}\mid _{\rm exact}
-\sigma^{(1)}\mid _{\rm app}\,,
\end{equation}
to exploit the fact that $\sigma^{(1)}\mid _{\rm exact}$ 
is known and $\sigma^{(1)}\mid _{\rm app}$ 
is included in $\sigma^{\rm res}$.
The difference between $\sigma^{\rm res}$ and $\sigma^{\rm
imp}$ is larger in $pp$ production, presumably because the $q \overline q$
approximation of $\sigma^{(1)}|_{\rm exact}$ is better than the $gg$ 
approximation.
We also show the NLO cross section calculated with the same mass and scale
factors as $\sigma^{\rm res}$ and $\sigma^{\rm imp}$.
The resummed and NNLO approximate
cross sections were calculated with the cut $s_4>s_0$ while no
cut was imposed on the NLO result. We can expect the resummation technique to
work well in both the $q \overline q$ and $gg$ channels
when $m/\sqrt{S} \leq 0.1$ because the NLO contribution is 
small compared to the LO term and the
perturbation series should then converge.  The effect of resummation is large
for $c \overline c$ production, increasing the cross section by a factor of
five or more relative to the NLO result.  
The agreement with the $c \overline c$ data from $\pi^- p$ 
\cite{Reu,Appel,pion} and $pp$ \cite{Reu,Appel,proton} interactions
is improved by the resummation.  The agreement with the existing $b \overline
b$ data from $\pi^- p$ \cite{bottom} interactions
is also somewhat improved.  In general, we see that $K_{\rm 
exp}^{\rm res}$ is much smaller than $K_{\rm exp}^{\rm NLO}$ for the same
$m$, $\mu$, and parton densities.
However, for charm quark production, the
technique is questionable above $\sqrt{S} = 15$ GeV.  The expansion parameter
is proportional to $\alpha_s(m^2) \ln(\sqrt{S}/m)$ so that when
$10\leq \sqrt{S}\leq 30$ GeV, $0.53 \leq \alpha_s(m_c^2) 
\ln(\sqrt{S}/m_c) \leq 0.83$ where $\alpha_s(m_c^2)$ for the GRV HO
distributions is found in Table 1.  We have shown the $c \overline c$ results
up to $\sqrt{S} = 30$ GeV even though the perturbative expansion no longer
converges and resummation fails in the $gg$ channel. This can be clearly seen
in the faster increase of $\sigma^{\rm res}$ and $\sigma^{\rm imp}$ with
energy compared to $\sigma^{\rm NLO}$ for $\sqrt{S}>20$ GeV in Fig.\ 3(a). 

We have also checked 
the resummation technique for a range of heavy quark masses, scales, and parton
densities. The variation in the results is indicated by the dotted curves
in Fig.\ 3.  Outside the range indicated by the dotted curves, the
resummation technique becomes questionable, particularly for the bound given
by the upper dotted curve.  When examining $b \overline b$ production, 
we studied the mass
and scale dependence over the range
$4.5 \leq m_b \leq 5$ GeV/$c^2$ and $m_b/2 \leq \mu \leq 2m_b$. The $b 
\overline b$ results are more stable
with respect to the variation, as indicated by the narrower bands in Fig.\ 3(c)
and (d).  We varied the charm quark mass between 1.3 and 1.8 GeV/$c^2$.  
Since the charm quark mass is relatively light, we study the scale
dependence indirectly. Note also that the resummation
technique cannot be applied to scales where $\mu < m_c$ because a perturbative 
treatment is uncertain for mass scales of $\sim 1$ GeV/$c^2$.

We also studied the parton density dependence so that along with the GRV HO
distributions we also used the GRV LO and the MRS D$-^\prime$ proton
distributions. Since the resummed
cross section depends on $\sigma^{(0)}$, it is instructive to
compare the results using LO parton densities with the one-loop coupling
constant to the NLO parton densities with the two-loop running coupling.
We also used the MRS D$-^\prime$ \cite{mrs} and SMRS P2 \cite{SMRS} pion
densities for $\overline{\rm MS}$ $\pi^- p$
production.  The low $x$ behavior of the SMRS densities is
different than the MRS D$-^\prime$
densities.  Presumably adjusting the behavior of the SMRS low $x$ region
to match that of the proton densities would shift the large $x$ contribution 
in the threshold
region.  Additionally, because $\Lambda_{\rm QCD}$ is different for the two
sets we use the value obtained in the proton fit.  Finally for
the MRS distributions the initial scale is larger than $m_c$, $Q_0^2 = 5$
(GeV$/c)^2$.  Therefore when we calculate $c \overline c$ production with 
these densities, we take $\mu = 2m_c$ instead of a full scale variation 
while we use $\mu=m_c$ for the GRV distributions.

The larger value of
$\alpha_s$ at one-loop both increases the cross section with the GRV LO
distributions as well as making the convergence of the perturbation series
slower for both $c \overline c$ and $b \overline b$ production.  In fact, since
$\alpha_s \sim 0.38$ at one-loop for $m_c = 1.5$ GeV/$c^2$, 
the resummation technique appears to fail
for the $gg$ case even at this relatively low energy, $\sqrt{S} = 15$ GeV.
In general, larger values of $\mu$ makes the convergence of the perturbation
series unreliable.  When $\mu = 2m$, $s_0$ is reduced by a factor of
four, resulting in a large $\overline \alpha_s$ in 
the exponent in eq.\ (6).  Thus a much larger $\mu_0$ 
is needed for convergence, $\mu_0 \sim 0.5m_c$ in the $q
\overline q$ channel and $\mu_0 \sim 0.7m_c$ in the $gg$ channel with the
MRS D$-^\prime$ distributions, much too
large to be compatible with the earlier results 
\cite{LSN,lsn2,NKJS,KS1}\footnote{We note that if the
MRS D$-^\prime$ distributions were refit with a smaller $Q_0$ so that $\mu =
m_c$ could be used, we would then expect the charm results to be similar to
those with the GRV HO distributions.}.
We also compared the $\overline{\rm MS}$ and DIS
schemes for the $q \overline q$ channel when appropriate.  For both the GRV LO
and MRS D$-^\prime$ DIS distributions (applied to $pp$ production only since
the SMRS pion distributions are only available in the $\overline{\rm MS}$
scheme) we
found convergence for a smaller value of $\mu_0$ and a correspondingly reduced
$\sigma_{q \overline q}^{\rm res}$, as expected.  However, since the gluon
can only be calculated in the $\overline{\rm MS}$ scheme for consistency
we give our results in the $\overline{\rm MS}$ scheme for both channels.

The extremes we found that still allowed the cross section to be resummed
are shown in the dotted curves of Fig.\ 3.  
Note that these calculations are not
consistent with the results obtained in the previous studies and are only meant
to indicate the possible range of $\sigma^{\rm res}$.  
The upper curves for $c \overline c$ production
are obtained with $m_c = 1.3$ GeV/$c^2$, $\mu = m_c$ and the GRV HO densities
while the
lower curves are calculated with $m_c = 1.8$ GeV/$c^2$, $\mu = 2m_c$ and 
the MRS
D$-^\prime$ densities.  For $b \overline b$ production, the upper curves are
calculated with $m_b = 4.5$ GeV/$c^2$, $\mu = m_b/2$ and the GRV LO 
densities.  The lower curves
are obtained with $m_b = 5$ GeV/$c^2$, $\mu = 2m_b$ and the 
GRV HO densities.  (Since $\mu > Q_0$ for the GRV HO and MRS D$-^\prime$
the same scales are used in each, producing very similar results.)
In Table 1 we show the appropriate value of $\alpha_s$ for each parton density
and scale used to calculate both our central results and our extreme cases.  To
illustrate how the resummation technique works at the extremes, in Figs.\ 4 and
5 we compare $\sigma^{\rm res}$ and $\sigma^{\rm app}$ for the upper and lower
bounds given by the dotted curves in Fig.\ 3.

In Fig.\ 4, the upper limit of
$c \overline c$ production obtained using the GRV HO distributions produces
convergence for $\mu_0 = 0.25m_c$ in the $q \overline q$ channel and $\mu_0 =
0.35m_c$ in the $gg$ channel.  These values,
with $m_c = 1.3$ GeV/$c^2$ are not so different from those we found with 
$m_c = 1.5$
GeV/$c^2$.  The smaller quark mass produces a stronger increase of 
$\sigma^{\rm res}$ with energy, as suggested in Fig.\ 3.  The $\pi p 
\rightarrow c \overline c$ cross section already diverges at $\sqrt{S} =
15$ GeV in Fig.\ 3(a).  The upper limit of $b \overline b$ production,
calculated with the GRV LO distributions, converges for $\mu_0 = 0.11m_b$
in the $q \overline q$ channel and 0.15$m_b$ in the $gg$ channel.  Note that
the smaller scale, $m_b/2$,
used for GRV LO results in a smaller $\mu_0$ needed for resummation.  However
the variation of both $\sigma^{\rm res}$ and $\sigma^{\rm app}$
with $\mu_0$ is quite strong since at $\mu_0 \approx 0.6m_b$,
$s_0/m^2 \approx 1$ and the separation of the $s_4$ integral is no longer
applicable.  In this case, the $b \overline b$ cross section begins to diverge
at large $\sqrt{S}$, due also 
in part to the large value of $\alpha_s$ for the lower scale.

In Fig.\ 5, the lower limit of
$c \overline c$ production obtained using the MRS D$-^\prime$ distributions 
produces a
convergent result for $\mu_0 = 0.5m_c$ in the $q \overline q$ channel and 
$\mu_0 = 0.7m_c$ in the $gg$ channel. The larger value of $m_c$ makes 
the resummation 
stable at higher energies.  However, the resulting $K_{\rm exp}^{\rm res}$ 
is so large this possibility appears to be ruled out.
The lower limit of $b \overline b$ production,
calculated with the GRV HO distributions, converges for $\mu_0 = 0.425m_b$
in the $q \overline q$ channel and 0.625$m_b$ in the $gg$ channel.  The ratio
$\mu_0/m$ is similar for the lower limits of $c \overline c$ production
since the scale is $\mu=2m$ in 
both cases even though different parton densities were used.  In general, it is
the value of the scale that determines the $\mu_0$ at which the resummed and
approximate results converge rather than the quark mass.  For example, in
$b \overline b$ production in $pp$ interactions, the MRS D$-^\prime$ and GRV
HO distributions give similar results for $\mu_0$ since $\mu = m_b$ can be
used for both \cite{KS1}.  Note that the 
higher mass and scale increases the relative $q \overline q$ contribution,
particularly for $\pi^- p$ interactions.

To summarize, we have presented a comparison of the NLO and resummed 
cross sections for 
charm and bottom production using similar $\mu_0$ values to those obtained
for top quark production.  The resummation of the S+V logarithms 
produces an enhancement of the NLO results.  The resummation technique
appears to work reasonably well
for $m_c = 1.5$ GeV/$c^2$ and $m_b = 4.75$ GeV/$c^2$, resulting also in 
$K_{\rm exp}^{\rm res} \approx 1$.  At the moment,
for a consistent description of both $c \overline c$
and $b \overline b$ production by pions and protons, 
the GRV HO distributions are the only choice available.  We reach this 
conclusion concerning the GRV HO densities for two reasons:  It is currently
the only set with the same low $x$ treatment of the pion and proton parton
distribution functions.  It also allows us to treat the scale on the same
level for all heavy quarks, $\mu = m_Q$, producing a uniform convergence
of the resummed cross sections at the same $\mu_0$ for charm, bottom, and
top.  Presumably another set of parton distributions that satisfies these
criteria would produce results similar to those given in Fig.\ 2 and the 
central curves of Fig.\ 3.  More precise
data on the charm, bottom and top production cross sections in the threshold
regions should clarify the situation and yield interesting information on the
interplay between perturbative and nonperturbative physics.\\

{\bf Acknowledgements}
The work in this paper was supported in part under the
contract NSF 93-09888.

%

\pagebreak

\begin{table}
\begin{center}
\begin{tabular}{|c|c|c|c|} \hline
\multicolumn{4}{|c|}{$c \overline c$} \\ \hline
& GRV HO (2 loop) & GRV HO (2 loop) & MRS D$-^\prime$ (2 loop)  \\ \hline
$\Lambda_3$ (MeV$/c$) & 248 & 248 & 280 \\ \hline
$m_c$ (GeV$/c^2$) & 1.3 & 1.5 & 1.8 \\ \hline
$\mu/m_c$  & 1 & 1 & 2 \\ \hline
$\alpha_s$ & 0.3 & 0.278 & 0.204 \\ \hline
\multicolumn{4}{|c|}{$b \overline b$} \\ \hline
& GRV LO (1 loop) & GRV HO (2 loop) & GRV HO (2 loop)  \\ \hline
$\Lambda_4$ (MeV$/c$) & 200 & 200 & 200 \\ \hline
$m_b$ (GeV$/c^2$) & 4.5 & 4.75 & 5.0 \\ \hline
$\mu/m_b$ & 0.5 & 1 & 2 \\ \hline
$\alpha_s$ & 0.312 & 0.187 & 0.155 \\ \hline
\end{tabular}
\end{center}
\caption[]{The values of $\Lambda_f$ and $\alpha_s$ for each of the parton
densities and scales we consider.  The left column gives the parameters for 
the upper
bound on $\sigma^{\rm res}$.  The middle column shows the parameters used
for our principle results.  The right column displays the parameters that give
the lower bound on $\sigma^{\rm res}$.}
\end{table}
\clearpage

\pagebreak
\centerline{\Large \bf Figure Captions}
\vspace{3mm}

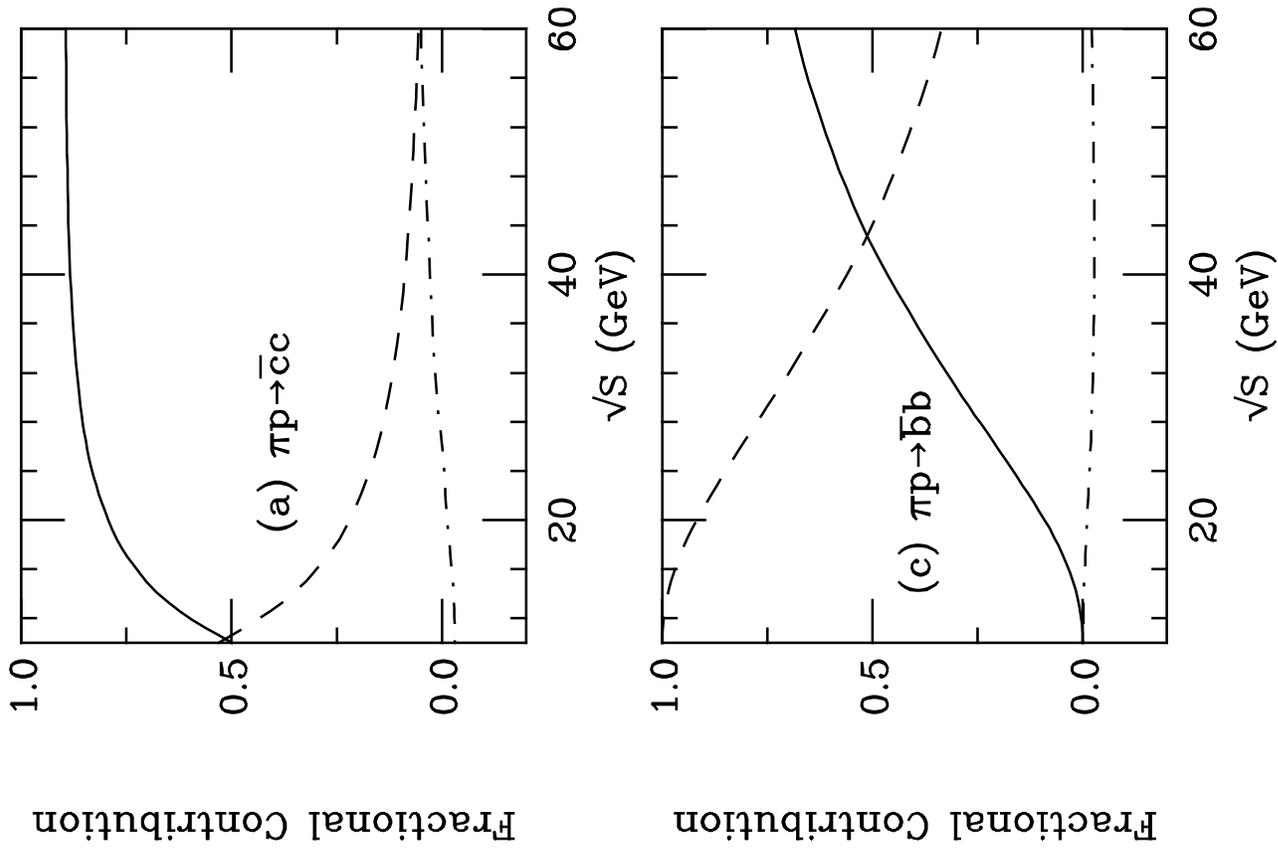
\noindent Fig. 1. Fractional contributions of the NLO channels to the
total $O(\alpha_s^3)$ $c \overline c$ and $b \overline b$ production cross
sections as functions of $\sqrt{S}$.  
We use $m_c = 1.5$ GeV$/c^2$, $m_b = 4.75$ GeV$/c^2$, 
and the GRV HO parton densities with the two-loop corrected
$\alpha_s$ from PDFLIB \cite{PDFLIB}.  All the
contributions are  given in the $\overline{\rm MS}$ scheme.
We show the $gg$ (solid), $q \overline q$ (dashed), and $(q+\overline q)g$ 
(dot-dashed), contributions to $c \overline c$ production
in (a) $\pi^- p$ interactions and (b) $pp$ interactions and $b \overline b$ 
production in (c) $\pi^- p$ and (d) $pp$ interactions.\\[1ex]

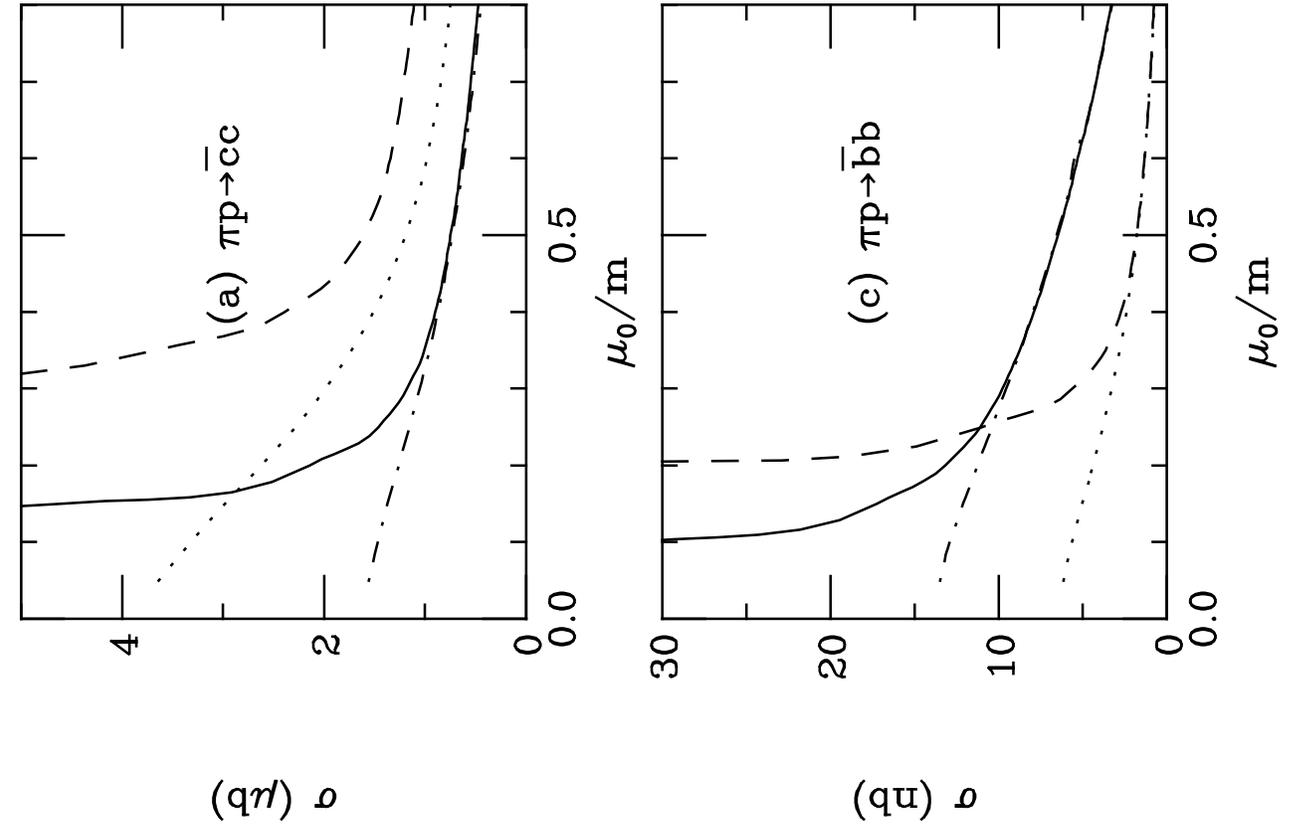
\noindent Fig. 2. The $\mu_0$ dependence of the resummed cross section for 
$c \overline c$ and $b \overline b$ production.  
We use $m_c = 1.5$ GeV$/c^2$, $m_b = 4.75$ GeV$/c^2$, 
and the GRV HO parton densities with the uncorrected two-loop $\alpha_s$.
All the contributions are  given in the $\overline{\rm MS}$ scheme.
We show $\sigma_{q\overline q}^{\rm res}$ (solid), $\sigma_{gg}^{\rm res}$ 
(dashed), $\sigma^{\rm app}_{q\overline q}$
(dot-dashed) and $\sigma^{\rm app}_{gg}$ (dotted)
 at $\sqrt{S} = 15$ GeV for
$c \overline c$ production
in (a) $\pi^- p$ and (b) $pp$ interactions and at $\sqrt{S} = 30$ GeV for 
$b \overline b$ production
in (c) $\pi^- p$ and (d) $pp$ interactions.\\[1ex]

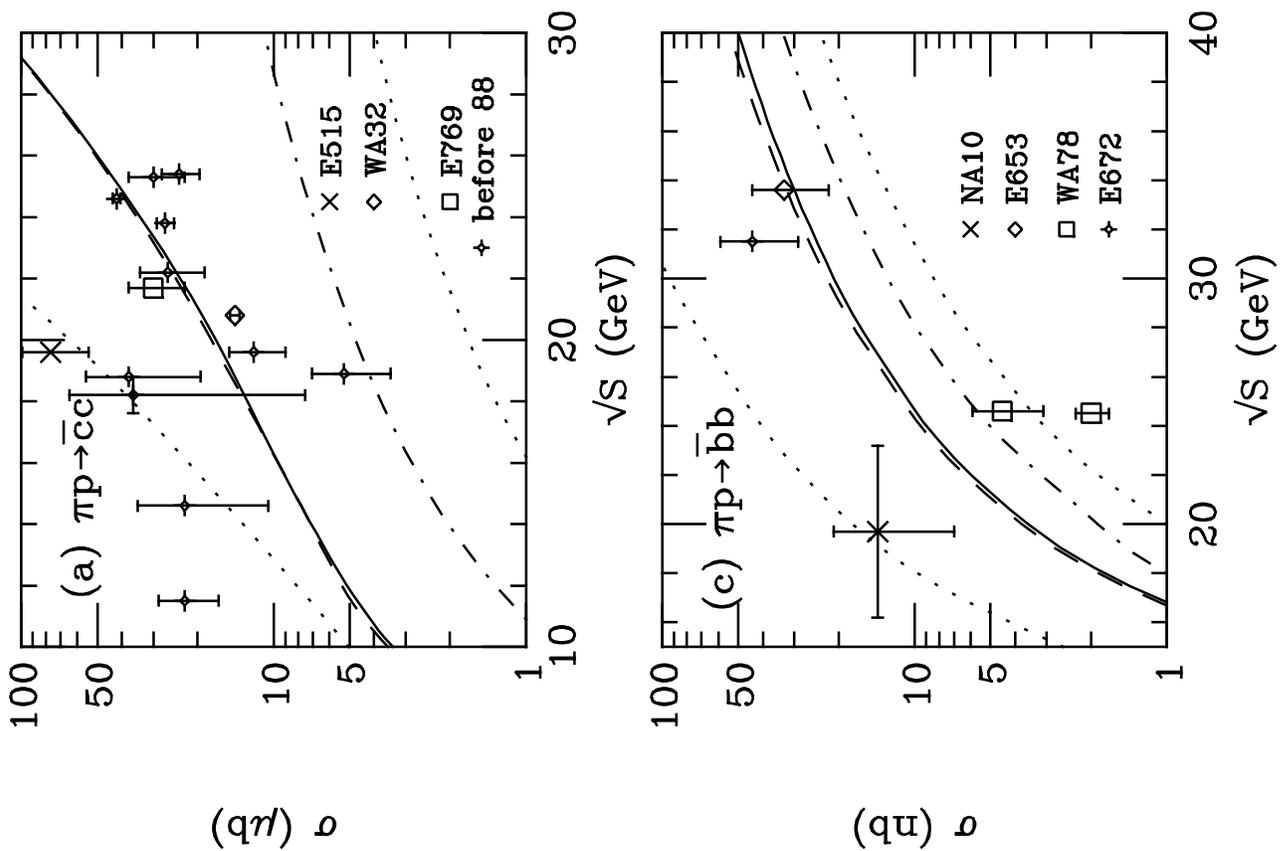
\noindent Fig. 3. The resummed, improved, and NLO cross sections 
are shown as functions of $\sqrt{S}$ for the
$c \overline c$ and $b \overline b$ production.
We show the total resummed cross section  
(solid) and the total improved cross section (dashed), both 
with $\mu_0=0.15m$ for the
$q \overline q$ channel and $\mu_0 = 0.35m$ for the $gg$ channel,
and the total $O(\alpha_s^3)$ cross section (dot-dashed) results
with $\mu=m$ using $m_c = 1.5$ GeV$/c^2$ and $m_b = 4.75$ GeV$/c^2$.  
We also show the extreme values of $\sigma^{\rm res}$ obtained when varying the
quark mass, scale, and parton densities in the dotted lines.
The results are given for $c \overline c$ production
in (a) $\pi^- p$ interactions and (b) $pp$ interactions and $b \overline b$ 
production in (c) $\pi^- p$ and (d) $pp$ interactions.\\[1ex]

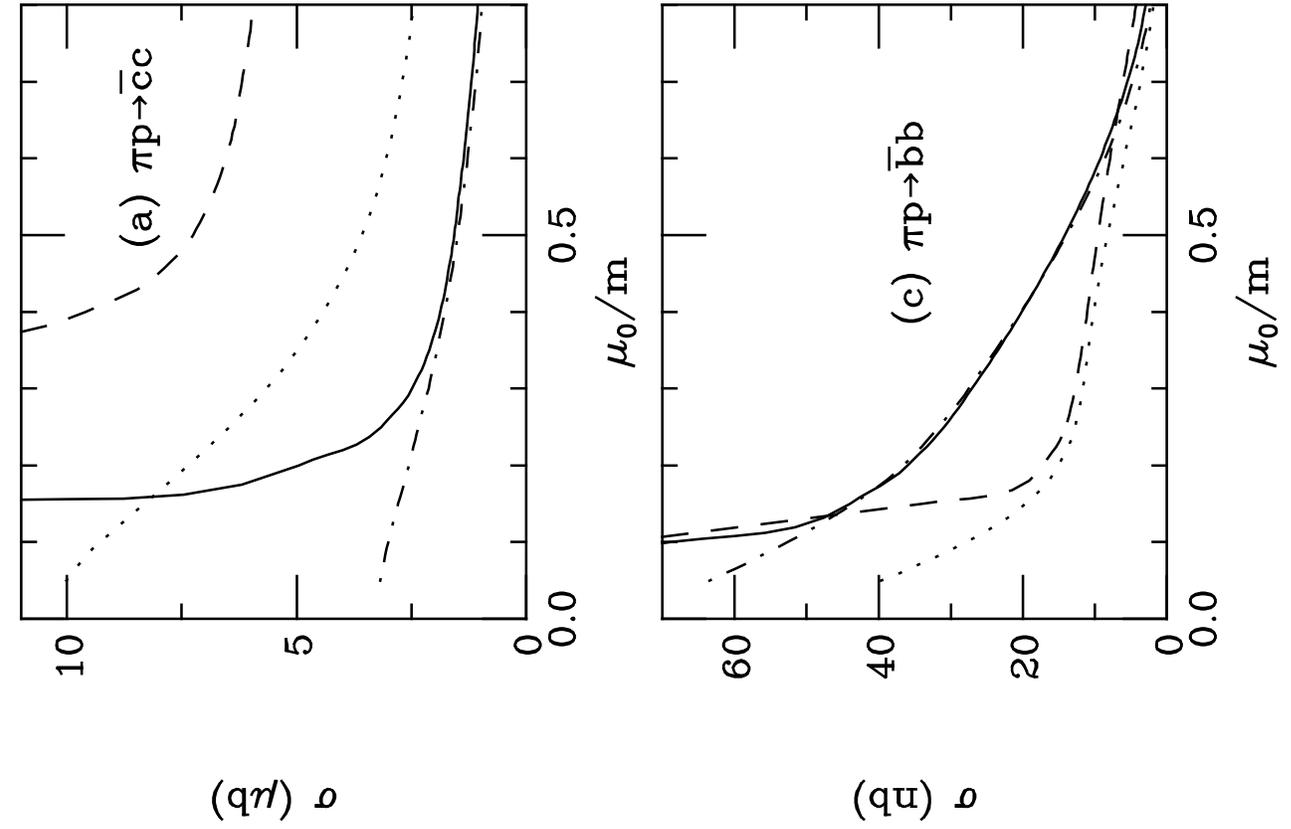
\noindent Fig. 4. The $\mu_0$ dependence of the upper bound of the
resummed $c \overline c$ and $b \overline b$ production cross sections
shown in the upper dotted curves of Fig.\ 3.  
We show $\sigma_{q\overline q}^{\rm res}$ (solid), $\sigma_{gg}^{\rm res}$ 
(dashed), $\sigma^{\rm app}_{q\overline q}$
(dot-dashed) and $\sigma^{\rm app}_{gg}$ (dotted) at $\sqrt{S} = 15$ GeV for
$c \overline c$ production
in (a) $\pi^- p$ and (b) $pp$ interactions and at $\sqrt{S} = 30$ GeV for 
$b \overline b$ production
in (c) $\pi^- p$ and (d) $pp$ interactions. For $c \overline c$
production, the curves correspond to the upper dotted curves in Fig.\ 
3(a) and (b) with $\mu = m_c = 1.3$ GeV/$c^2$ and the GRV HO densities.
For $b \overline b$
production, the curves correspond to the upper dotted curves in Fig.\ 
3(c) and (d) with $2\mu = m_b = 4.5$ GeV/$c^2$ and the GRV LO densities.\\[1ex]

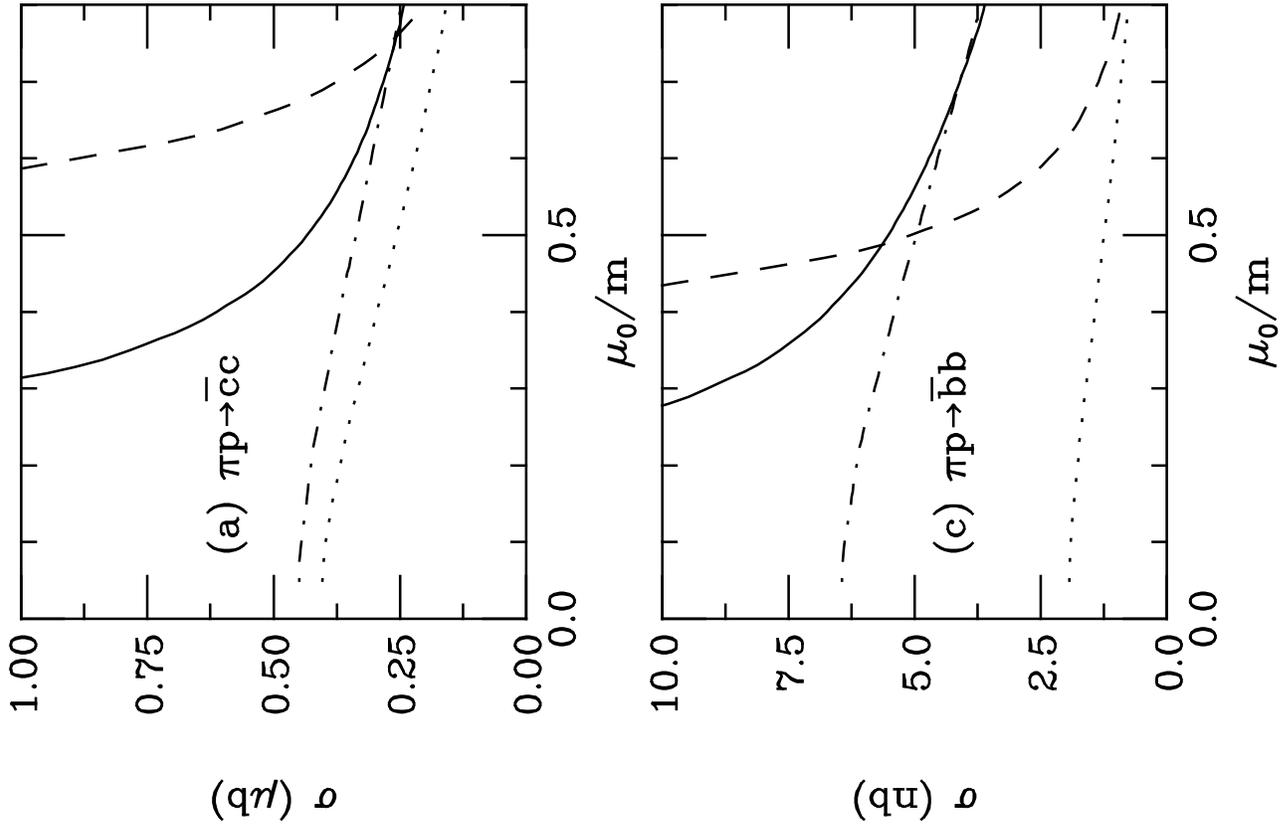
\noindent Fig. 5. The $\mu_0$ dependence of the lower bound of the
resummed $c \overline c$ and $b \overline b$ production cross sections
shown in the lower dotted curves of Fig.\ 3.  
We show $\sigma_{q\overline q}^{\rm res}$ (solid), $\sigma_{gg}^{\rm res}$ 
(dashed), $\sigma^{\rm app}_{q\overline q}$
(dot-dashed) and $\sigma^{\rm app}_{gg}$ (dotted) at $\sqrt{S} = 15$ GeV for
$c \overline c$ production
in (a) $\pi^- p$ and (b) $pp$ interactions and at $\sqrt{S} = 30$ GeV for 
$b \overline b$ production
in (c) $\pi^- p$ and (d) $pp$ interactions. For $c \overline c$
production, the curves correspond to the lower dotted curves in Fig.\ 
3(a) and (b) with $\mu/2 = m_c = 1.8$ GeV/$c^2$ and the MRS D$-^\prime$ 
densities.  For $b \overline b$
production, the curves correspond to the lower dotted curves in Fig.\ 
3(c) and (d) with $\mu/2 = m_b = 5$ GeV/$c^2$ and the GRV HO densities.\\[1ex]

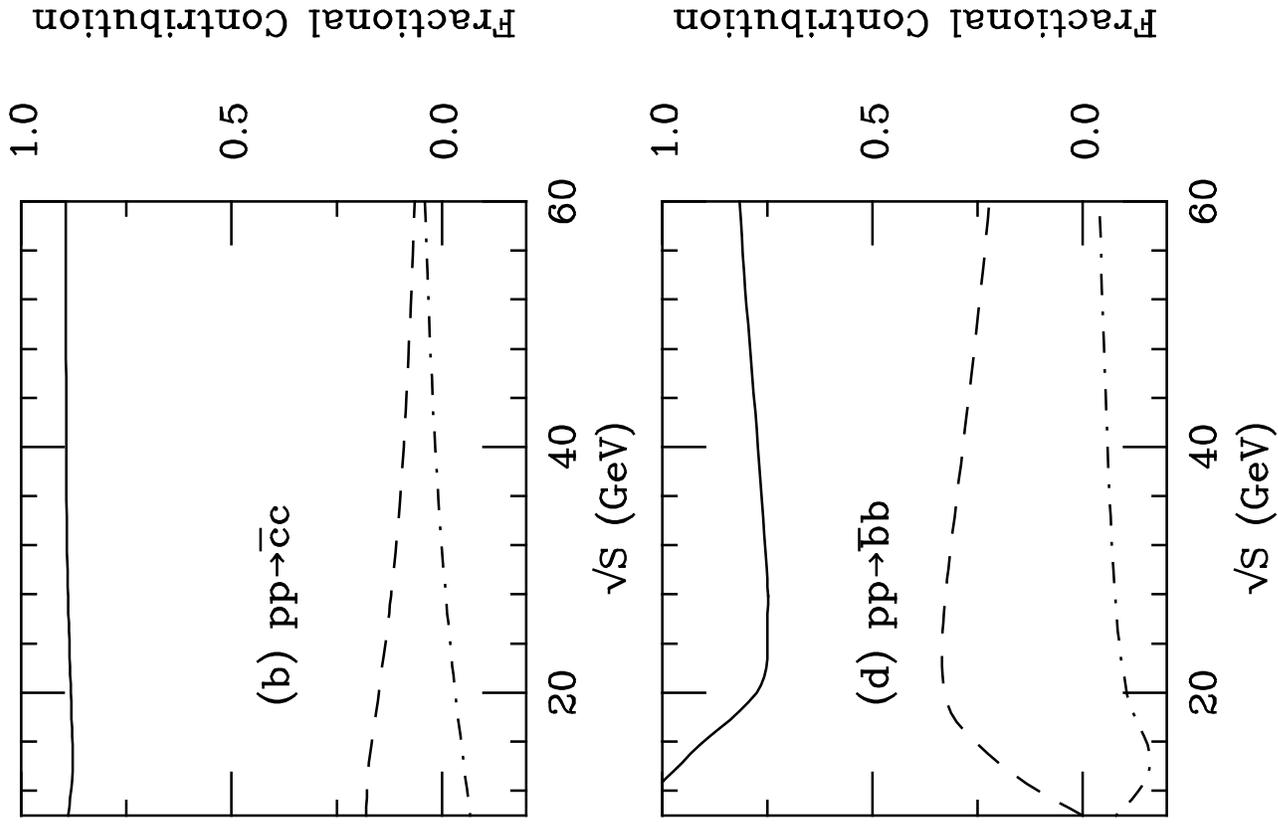
\end{document}